# Free Will: A New Formulation

Eric Sanchis
University of Toulouse Capitole
Toulouse, France
e-mail: eric.sanchis@iut-rodez.fr

*Abstract*: **Free will is sometimes summarised in the philosophical literature as the subjective impression felt by an individual that he or she is the ultimate source or cause of his or her own choices. The two most common arguments for denying the existence of free will come from philosophy and neuroscience. The first argument is the Consequence Argument. The second asserts that our decisions are first made by the brain and only then become conscious to the subject, taking away the control of the decision. The purpose of these two arguments is to demonstrate that an individual cannot be the source or primary cause of his or her choices. It is shown in this work that the concepts of primary cause and primary source are not adequate to state a solid characterisation of free will. A new formulation of this property is proposed in which it is seen as a three-stage decision-making process implemented by an individual to escape his or her own real or supposed alienation. This decision-making process is represented in the form of a computer model called the PSU (Predictability - Suspension - Unpredictability) model. The compatibility of this new formulation of free will with the feeling it provides and the analysis of various situations are then discussed.**

*Keywords*: **Free will, alienation, first-person indeterminism, three-stage decision-making process, computer model.**

## 1. Introduction

The questioning of free will, whether on its interpretation or its existence, continues to fuel a centuries-old debate that the philosopher Peter van Inwagen has succinctly summarized as follows: "No philosopher has achieved an understanding of free will. That may be because free will is indeed something that human beings are incapable of understanding, but it may be because we human beings have not yet discovered the right way to think about free will" [12].

It is true that free will is a real complex system as it intertwines vague concepts and relationships that are difficult to define [7]. Essentially studied in philosophy, the contemporary debate on free will has been enriched in recent decades by new contributions from disciplinary fields such as neuroscience and psychology, fuelling the original controversies without definitively settling the central question about free will: does free will exist?

To answer this difficult question, a multi-stage approach was adopted. The first step was to start with a description of the free will feeling experienced by a person as it is generally formulated: free will is the subjective feeling felt by an individual that he is the source and ultimate cause of his own choices. Hereafter, this feeling of free will will be called *internal viewpoint* or *first-person free will*. This *internal viewpoint* will then be confronted with the discourses and results of two fields that are particularly critical of the existence of free will: philosophy (based on the results provided by the physical sciences) and neuroscience. The views from philosophy and neuroscience will be described as *external*. It will be shown that the concepts of *first cause* or *ultimate source* used in the *external views* are too fragile concepts to build a solid characterisation of free will.

In the second part, the consequences of this fragility are drawn: it is necessary to reformulate the conditions of existence of free will since the concepts of first causes and sources are not sufficiently solid. Free will is now considered as a decision-making process implemented by an individual in order to escape a possible alienation, i.e., a possible conditioning. This second part ends with the description of a model, called the (*Predictability - Suspension - Unpredictability*)

*PSU model*, in which three phases follow each other in a precise order in a way to achieve this objective.

In the last part, the compatibility of the *PSU model* with the feeling of free will and the legitimacy of using randomness are discussed. The model is then used to analyse situations in which free will has been invoked to deny its existence.

## 2. Free Will and Its Problems

The feeling of free will gives the person who experiences it a sense of personal freedom in relation to his or her own actions. In order for this freedom to be exercised correctly and for this feeling to take place, several conditions are generally laid down. The subject must have several possible effective choices (*alternative possibilities*), i.e., several choices that are actually feasible. Furthermore, in order to make a decision, the subject must be able to decide without external pressure or constraint. This situation is summarised in the philosophical literature as follows: the subject with free will is the primary source (or cause) of his choice. This *internal view* of free will constitutes a *first-person free will* because it is felt directly by the subject.

### 2.1 Universal determinism

*1) Third-person free will:*

In contrast to the *internal viewpoint*, the *external viewpoint* studies the possibilities of the existence or non-existence of free will by considering the laws of physics that govern Nature (and therefore any individual). We will call these laws the *laws of nature*. There is no consensus on the content of this expression. Some consider these laws to be "immutable and objective" [11], others associate them with the different laws discovered in the different fields of knowledge (e.g. physics, chemistry, biology, etc.). For our part, we will identify the *laws of nature* with the most fundamental physical laws of the universe as they are known to us at the present time. They can therefore change according to the state of our knowledge.

The *external viewpoint* defines a *third-person free will* because it corresponds to a scientific view of the world. It is based on the concept of *determinism*. There are different meanings and names for this concept: methodological determinism, causal determinism, nomological determinism, universal determinism, local determinism, concrete determinism and many others. One can distinguish between 'strong' versions of determinism (nomological determinism, universal determinism) and 'weak' versions (local determinism, concrete determinism) [8][5].

The existence of free will is usually confronted with a 'strong' version of determinism. The version that will be used is *universal determinism*. *Universal determinism* combines *physical determinism* and the principle of *universal causality*. *Physical determinism* is the thesis that there is at all times a single possible physical future. This is fixed by the past and by the *laws of nature*. The principle of *universal causality* states that every event in the universe has a cause.

The acute problem with *universal determinism* is that it conflicts head-on with the *internal view* of free will. A refutation of the existence of free will by *universal determinism* is called the *Consequence Argument* [11]. This reasoning can be presented as follows. According to the *universal determinism* thesis, the present is caused by the conjunction of the *laws of nature* and events that occurred before we were born. As a result, the choice we make in the present moment is the product of a long sequence of causes and effects that are beyond our control. The choice we thought was ours was in fact already fixed by the causal chain: we are therefore not the ultimate source of our choices or of any of our actions. Therefore, free will is an illusion [4]. This opposition between determinism and the existence of free will is called the *Free Will Problem*.

*2) The primary source problem:*

According to the view expressed in this work, the weakness of *universal determinism* in relation to the existence of free will stems mainly from the principle of *universal causality* because it is possible that this principle does not apply to all moments in the universe.

The principle of *universal causality* states that every event in the universe has a cause. The first problem with this principle is that it logically leads to a regression to infinity.

The second problem comes from physics. Today, the physical theory associated with the evolution of the universe is the *Big Bang theory*. According to this theory, when we go back in time to what is called the *singularity*, we can no longer refer to an *instant 0* or a *first cause*: these two notions vanish because the mathematical equations no longer have any physical meaning. In other words, this means that we can say nothing about the existence (or non-existence) of an instant 0 or a *first* or *ultimate cause*. It means that this *instant 0* or this *first cause* can exist as well as not exist.

Therefore, defining free will as the capacity of a subject to be the primary source or the primary cause of its choices becomes problematic.

### 2.2 Free will and Neuroscience

*1) Libet's experiments:*

The second area in which free will has been carefully studied is neuroscience and its satellite components such as neuropsychology. Libet's pioneering experiments in the early 1980s [9] added to the debate on free will by showing that a decision is made in advance by the brain before it becomes conscious to the subject. These experiments consisted of asking a subject to watch a clock of some sort ticking away and, when he decided to do so, to rapidly flex the fingers or wrist of his right hand. He was also asked to identify the moment when he made his decision. The subject was equipped to distinguish between three moments: the onset of neural activity, the moment of the subject's decision and the onset of the actual movement. These experiments showed that brain activity began several hundred milliseconds before the subject's decision making became conscious. Various experiments of this type have been conducted and have led to this same result [1]. These experiments suggested or supported the idea that free will could not exist since it was our brain that decided for us, with awareness of the decision coming much later. Since then, new experiments have shown that the time lag between preparatory brain activity and awareness of the action could be much shorter, or even non-existent, leaving the question of free will open [10].

*2) The problem of the beginning of conscious action:*

Libet's experiments raise the thorny problem of the moment when the conscious action begins. Indeed, it is possible to consider that this moment corresponds to the moment when the subject decides to raise his wrist or well before, when the subject has understood the instruction. Although in both cases the decision to perform the movement may become conscious after preparatory brain activity, the overall interpretation of the experiment may be quite different.

To illustrate this, let us place ourselves in the very general framework where the functioning of the human cognitive system is considered to be the result of continuous interactions between a conscious and a non-conscious system, with the two systems influencing each other [Fig. 1].

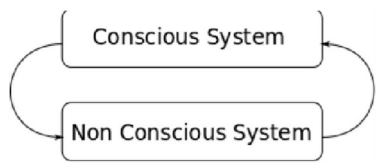

Fig. 1: Human cognitive system general functioning

One interpretation of Libet's experiments could be the following. After the subject has consciously assimilated the instruction, his non-conscious system detects a situation that has already been experienced many times (choosing an element from a set where all the elements of this set have the same value), that there is no need to mobilise the subject's high-level cognitive functions such as reasoning or introspection to carry out the requested task and that an automatic response provided by the low-level cognitive functions is sufficient. Using Kahneman's terminology [6], the situation detected by the *system level 1* (called *S1*) leads the latter to make some kind of decision (raise the wrist at time *t* and not at time *t'*), which does not require any justification, but which respects the instruction. The major interest of *S1*, a component of the non-conscious system, is that it operates very quickly, automatically and at a low energy cost [2].

From the point of view of the subject's high-level cognitive system, *S1* behaves like a random generator providing him/her with simply one answer from a set of possible answers, each of which is of equal importance since it is meaningless. In other words, the subject has unconsciously delegated to his *S1* system the making of a random decision which the subject will interpret as his own.

We can see that an individual has two very different ways of obtaining an unpredictable response: the explicit use of an external device such as a die or the implicit use of his or her *S1* system. In the first case, the triggering of the device is conscious, in the second case it is unconscious. Nevertheless, in both cases, the result is the obtaining of a response that is not predictable for the subject, i.e., the implementation of a *first-person indeterminism*. It should be noted that the 'quality' of the random generator used is of little importance to the individual: only the non-predictability is important.

The following section will show how this *first-person indeterminism* is an essential component of free will. To this end, it will be necessary to clarify how randomness and free will are related. Indeed, these two notions are often put in opposition, a random decision being generally posed as the opposite of a carefully chosen decision.

## 3. Free Will as a Three-Stage Decision-Making Process

### 3.1 A reformulation of the free will property

It was established earlier that the characterisation of free will incorporates two quite distinct elements: the feeling it provides and the conditions required for it to exist. One of these conditions requires that the subject must be the primary cause of his choice. However, the notion of *cause* (and a fortiori that of *first cause*) corresponds in most cases to a hypothetical interpretation of a situation, i.e., to a fiction.

In order to avoid the unstable concept of *cause*, the property of free will will be reformulated using the notion of *alienation* and *conditioning*. More precisely, free will will be considered as a decision-making process implemented by a subject to extract himself from a possible alienation, i.e., from a dispossession of his own control. Indeed, because of his cognitive and social constitution, an individual can never be sure that the decision he has just taken is not the product of a conscious or unconscious conditioning, the result of internal (e.g. habits) or external (e.g. education, social practices) constraints.

More precisely, we will say that an individual uses his free will when, after having made a decision *C* in a given context, i.e., in which the subject has a precise choice for each of the situations associated with this context, he decides to suspend this choice and to make a random choice *C'*. In other words, the subject changes his mind by replacing a predictable choice with a non-predictable choice. This unpredictability of the final choice is a *first-person indeterminacy* directly experienced by the individual.

Let us consider the following example which will serve as a guideline. Let us suppose that an agent *A* has to go the next day to a point *P* some ten kilometres away. As usual, agent *A* decides to adopt the following behaviour: if it is raining, he will use his car, if it is not raining and the sky is grey, he will ride his bicycle. Finally, if the weather is good, he will walk to point *P*. The next day, the sun shines. Logically, agent *A* should walk, but being introspective, he asks himself: why did he associate good weather with walking? Is it because his doctor has told him that he is not getting enough exercise? And if it were raining, why would he have to take the car though he prefers cycling? Is it because his education taught him that rain causes accidents and colds?

In this example, agent *A* will use his free will by suspending the pre-determined choice, i.e. going to point *P* by walking, and randomly choosing one of the following four possibilities: using one of the three modes of locomotion (walking, cycling, car) and choosing not to go to point *P*. In this decision-making process, the existence or not of physical determinism (or indeterminism) is irrelevant. What matters for agent *A* is his ability to question his possible alienation and to execute a possibly different choice decoupled from the current situation.

### 3.2 The PSU Model

The decision-making process related to free will as described above has three phases: a first phase where the outcome is predictable, a phase where the execution of the predicted choice is suspended and a last phase where the final choice is not predictable by the subject because it is drawn at random. This process can be modelled by the sequential execution of two components: a predictable component and a non-predictable component [Fig. 2].

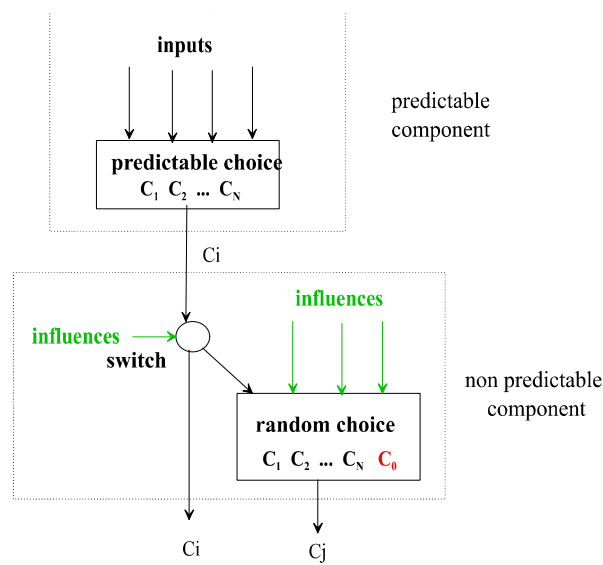

Fig. 2: PSU Model

*1) Predictable component:*
The first component that runs has a predictable behaviour. This can be represented in the form of a set of parameters which will be called *inputs* and which directly drive the operation of the component. At the end of the execution of the predictable component, one of several choices is made.

From a formal point of view, the predictable component is defined by a triplet:
- a set of choices $C_p$ ($1 \leq p \leq N$, $N > 1$),
- a *selection function* and
- a set of *inputs*.

From the value of the *inputs* and the *selection function*, the predictable component makes a choice $C_i$ from the $N$ possible choices.

The component is said to be predictable because if at two different times $t$ and $t'$, the same *inputs* are applied to it, then the component will produce the same behaviour: the *selection function* will produce the same choice.

Finally, the notion of *input* designates a well identified information having an effect on the *selection function*. It can represent an internal or external condition, a stimulus, a state or any other type of known information.

In this formal framework, the previous example can be modelled as follows:
- the set of choices includes three possibilities: take the car, take the bike, walk
- a single *input* controls the *selection function*: the state of the weather.

In the described situation, the weather is good: therefore, the *selection function* of agent $A$ will choose walking.

*2) Unpredictable component:*

The second component that is executed implements a non-predictable operation. It comprises two very distinct modules:
- a *switch module*, which can have two possible effects on the choice $C_i$ produced by the previous component: a *transparent* behaviour and a *modifying* behaviour. In the *transparent* mode, the choice $C_i$ resulting from the predictable component will finally be executed. In the *modifier* mode, the previous choice $C_i$ is temporarily suspended and a *non-predictable selection function* is activated
- a module consisting of a *non-predictable selection function*. When this selection function is activated, it makes a random choice $C_j$ out of $N+1$ possible choices: the $N$ choices coming from the predictable component to which is added an additional choice noted $C_0$. This choice models the final inhibition of the previous choice $C_i$, by the non-predictable component. The meaning of the choice $C_0$ depends on the context (veto, stop of the activity, other).

The interest of using chance is that it is neutral because it cannot be influenced.

As a result, after execution of the non-predictable component, the choice $C_j$ ($0 \leq j \leq N$) finally selected may be different from the choice $C_i$ made by the predictable component. The final choice depends essentially on the operating mode of the *switch* (*transparent* mode or *modifying* mode).

It should be noted that the way in which the *switch*, and more broadly the non-predictable component, operates depends on internal conditions that are not necessarily known to the agent implementing this decision process. These internal conditions which are partially or totally unknown and which play a role similar to that of *inputs* for the predictable component are called *influences*.

It is now possible to illustrate how agent $A$ in the previous example implements its free will. As a reminder, at the time agent $A$ has to go to point $P$, the weather is fine. Therefore, agent $A$'s predictable *selection function* prescribes that he should walk to his destination (choice $C_i$). But a doubt arises in $A$'s mind: "Am I the source of this choice or is it the product of some alienation (my education, my prejudices, etc.)? This doubt can be the result of the functioning of the agent's *S1* system (non-conscious system), of a remark made by those around him or her, or of any other event. Agent $A$ then decides to question this choice $C_i$ (switch to *modifier* mode) by drawing one of four choices: the three previous choices (go to point $P$ by car, by bicycle, by walking) and the choice not to go to point $P$ (choice $C_0$), i.e., to inhibit the action.

Agent $A$ can make this non-predictable choice by using, for example, an appropriate computer program or any other device that allows him to draw a value among four. It should be noted that for the user the quality of the random choice function is of little importance in the overall decision-making process.

In this particular example, the notion of *influence* allows for the representation of both an intuition and a change in the mood of agent $A$ that caused the switch to be switched to *modifier* mode. Its major interest is that it avoids having recourse to the concept of *cause*, a cause that could only be speculative in this context.

# 4. Discussion

### 4.1 Analysis of the Model

*1) Feeling and conditions of existence of free will:*

It was established that it was important to distinguish the feeling of free will from the conditions of existence of this property. In particular, it was shown that conditioning the existence of free will on the possibility of a subject being the primary cause of his choice meant relying on two fragile concepts when applied to an individual: the notions of *cause* and *source*. As this characterisation was deemed unsatisfactory, it was proposed to consider free will as the capacity of a subject to escape from a possible conditioning, whether internal or external.

If this reformulation of free will is perfectly compatible with the feeling that this property should provide, the question remains whether the decision-making process proposed to implement it is also compatible with this feeling. Indeed, using randomness to make a decision is generally considered to be contrary to the exercise of one's free will. Yet this negative understanding of the use of randomness is inconsistent with certain practices in human societies, such as the drawing of lots for political representatives, juries and the like. Would our societies behave irrationally? The answer is obviously no. It was after analysing the negative effects of previous practices that this voting system was introduced and is still used today because chance is neutral and uninfluential. In a similar way, in the *PSU model*, the drawing of lots by the agent follows an interrogation of past experience in order to correct any negative effects.

*2) From the model to practice:*

Having defined the *PSU model*, the question to consider is whether this decision-making process is actually used in practice. It is clear that few people use a die or other external device to experience this sense of free will.

The model as presented is in fact a *canonical model*, i.e., it describes the ideal implementation of the free will property. In everyday life, this decision-making process is most often simplified: after the suspension of choice $C_i$, the agent does not usually use an external random device (die, computer program) but rather his non-conscious system (*S1 system*). It is the latter that suggests a new choice $C_j$. This mode of operation has several advantages. First of all, the choice elaborated by the non-conscious system is not predictable for the conscious system. Secondly, for the agent, the use of its non-conscious system is faster and, above all, much less energy consuming. Finally, the availability of an external random device is not always guaranteed. On the other hand, the major disadvantage of this degraded mode of operation is that it does not provide any guarantee with respect to the objective set: to break the possible alienation of the agent. Indeed, it is known that the non-conscious system is the receptacle of the conditioning undergone or produced by the agent. Nevertheless, the execution of the *canonical model* is possible by any individual who wishes to do so.

### 4.2 Using the Model

The characteristics of the decision-making process described in the *PSU model* open up interesting perspectives that can easily be imagined, whether in cognitive computing or in the creation of advanced artificial entities (games, human-computer interfaces, societies of artificial agents, etc.). In a more original way, this model can be used to analyse real-life situations (attitude of participants in Libet's experiments) or literary situations (attitude of the Lafcadio's character).

*1) Libet's experiments:*

Libet's experiments have sometimes been invoked to justify the non-existence of free will. The three-phase model of free will allows us to analyse these experiments without going into the details of the experiments. According to this model, these experiments have little to do with free will. Indeed, the first observation is that the subjects are under influence: they are asked to perform one task and cannot do another. They are not in a usual situation for which they would have different strategies and for which they would like to escape from possible conditioning. In particular, they do not have the possibility to abstain, i.e., to express a veto right.

*2) Lafcadio and the act without motive:*

In his novel **The Vatican Cellars**, André Gide [3] describes a scene in which a young man named Lafcadio travels in the same train compartment as an elderly person named Fleurissoire. They are alone in the compartment and Fleurissoire is standing by the door. On a whim, Lafcadio decides to gamble Fleurissoire's life on the occurrence of a random event. He will count slowly to 12: if he perceives the presence of a fire in the countryside before the end of the count, then he will push Fleurissoire out of the train, if not, he will refrain and Fleurissoire will live. When he reaches 10, Lafcadio sees a fire and commits his crime.

This scene is interesting for several reasons. Firstly, Lafcadio wants to prove to himself that he is capable of performing an act without motive. Secondly, he creates an external random device to make his final decision (appearance of a particular event after a finite time).

The act without a motive immediately refers to the act without a cause. However, as was pointed out earlier, it is extremely difficult to univocally associate a causal chain with a human act, and even more difficult to demonstrate that this causal chain is empty. Moreover, a causal chain must have an origin, i.e., an initial cause, etc. We are confronted with an artificial and useless regression to infinity which does not provide any definitive concrete explanation.

Lafcadio's random external decision making device might seem to correspond to the previous three-phase decision making process. However, this is not the case. Using the terminology of the model, the decision-making process performed by Lafcadio begins with a sudden impulse that could be assimilated (incorrectly) to the *switch* in the model, followed by the execution of an external component providing a random outcome. In fact, the switch is not really a switch since it has only one mode of operation. Indeed, Lafcadio's sudden impulse is not preceded by any rational choice: Lafcadio is not, for example, a professional killer with his own preferences (this mode of killing for this type of individual, that mode of killing for another, etc.). Lafcadio does not try to escape any conditioning related to a professional habit. Indeed, he shows that he is unable to explain his act when he says: "How do you expect me to explain to you what I cannot explain to myself? ".

## 5. Conclusion

Associating free will with the concept of *alienation* of an individual has the advantage of no longer appealing to a notion as delicate as that of *first cause*. The existence or non-existence of a physical or ontological determinism loses its importance because the agent is satisfied with an unpredictability that he creates himself.

By distinguishing the *canonical model* of free will from its empirical use, a better understanding of this property has been possible. Free will is no longer a mere illusion because exercising free will corresponds to the execution of a specific decision process in a specific context.

Finally, in addition to Artificial Intelligence which is the natural domain of the *PSU model*, the field of application of this model could be extended to the analysis of technical aspects developed in philosophy of mind, such as the Manipulation Argument.